\documentclass{article}
\usepackage{amsfonts,amssymb, amsmath}
\usepackage[english]{babel}
\usepackage{graphicx}
\usepackage{float}

\newtheorem{prop}{Proposition}

\newtheorem{ex}{Example}
\newenvironment{exam}{\begin{ex} \rm }{\end{ex}}

\def\bq{ \begin{equation}}
\def\eq{ \end{equation}}
\def\ben{ \begin{eqnarray}}
\def\en{ \end{eqnarray}}

\begin{document}


\title{Discretization and superintegrability  all rolled into one}

\author{A.V. Tsiganov \\
\it\small St. Petersburg State University, St. Petersburg, Russia\\
\it\small e--mail: andrey.tsiganov@gmail.com}
\date{}
\maketitle

\begin{abstract}
Abelian integrals appear  in  mathematical descriptions of various physical processes.  According to Abel's theorem
these integrals are related to  motion of a set of points along a plane curve around fixed points,
which are rarely used in physical applications. We propose to interpret coordinates of the fixed points either
as parameters of exact discretization or as additional first integrals for equations of motion reduced to Abelian quadratures on a symmetric product of algebraic curve.

 \end{abstract}

\section{Introduction}
\setcounter{equation}{0}
Most of the nowaday applications of Abel’s theorem use  Riemannian ideas and, therefore,  in current textbooks Abel's Theorem looks as follows:
\par
\textbf{Modern version of Abel's theorem:} \textit{
Let $X$ be a compact Riemann surface and $D$ be a divisor of degree zero on $X$. Then $D$ is the divisor of a
meromorphic function on $X$ if and only if $\mu(D)$=0 in the Jacobian of $X$}.
\par\noindent
Here $\mu:$  Div$X\to$Jac$X$ is an Abel-Jacobi map, so if $\mu(D)=0$, there is a  collection of paths $\lambda_k$ from   base point $P_0\in X$ to points $P_k$ in the divisor $D$ so that
\[
\sum_k \int_{\lambda_{k} } \omega=0\,.
\]
This  theorem has well-known roots in classical mechanics.  Indeed, in 1694  James Bernoulli studied the curve for which time taken by an object sliding without friction in a uniform gravity to its lowest point is independent of its starting point, and introduced integrals which can not be expressed in terms of elementary functions.  Similar integrals were discovered in attempts to rectify elliptical orbits of planets, so such integrals became known as “elliptic integrals.”  Later,  Euler and Lagrange provided an analytical solution to the so-called tautochrone problem and  applied the addition law for elliptic integrals to search of  the algebraic trajectories among transcendental ones in the two fixed center problem \cite{eul1,lag}.

Therefore, it is not surprising that  Abel in his M\'{e}moire \cite{ab}  studies integrals of algebraic functions using rational time parametrisation of a plane curve and motion of  variable points along this curve, see historical remarks in \cite{grif04}.
\par
\textbf{Original version of Abel's theorem:}
 \textit {
 A set of $k$ points $(x_j,y_j)$ moving along plane curve $X$ can be subjected to a finite number of algebraic constraints in such a way that a sum of indefinite integrals
\bq\label{abel-sum}
\int \omega(x_1,y_1) dx_1+\int\omega(x_2,y_2) dx_2+\cdots+\int\omega(x_k,y_k) dx_k
\eq
can be expressed in terms of algebraic and logarithmic functions of  coordinates $(x_j,y_j)$ of the moving points provided these coordinates satisfy the constraints.}
\par
Algebraic constraints are independent of differentials $\omega(x,y)dx$ and, according to Clebsch and Gordan, we can replace the "algebraic constraints" with "coordinates of fixed points". Movable and fixed points form a divisor which division into two types of points allows to describe the so-called canonical injections of $k$-fold symmetric products $X(k)$  of  algebraic curve $X$
\[
j_{mk}:\qquad X(k)\to X(m)\,,\qquad k>m\,,
\]
which are compatible with the  Abel-Jacobi  map $\mu$.   In \cite{chow}  Chow proposed a projective construction of the Jacobian using these injections and the Riemann theorem. This exhibited  basic character of the Jacobian in a new way. It was taken up by Matsusaka and later by Grothendieck in their works on the Picard variety, see discussion in textbook \cite{gun}.

\begin{exam}
As an illustration of this generic theory we  take cubic curve $X$  defined by a short  Weierstrass equation
\bq\label{wei-eq}
 X:\qquad y^2=x^3+ax+b\,.
\eq
and  consider variable points of intersection of $X$  with a family of
straight lines  all passing through the same fixed point and depending  on  parameter $t$
 \[Y:\qquad y=b_1(t) x+b_0(t)\,.\]
 \begin{figure}[!ht]
\center{\includegraphics[width=0.7\linewidth, height=0.2\textheight]{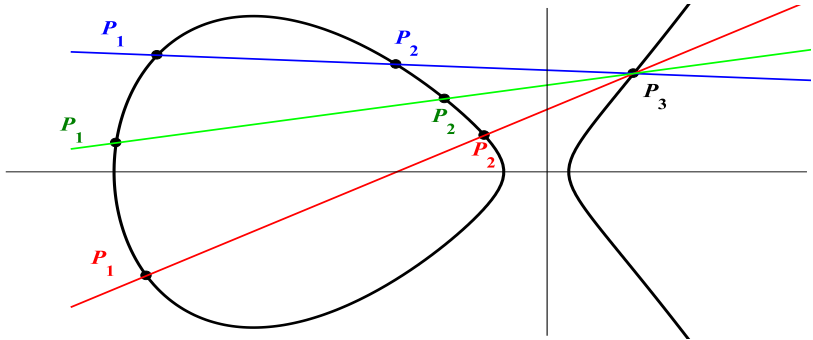} }
\caption{Motion of points $P_1$ and $P_2$ around fixed point $P_3$ on $X$}
\end{figure}
\par\noindent
 In his proof of  Euler's results Lagrange identified $t$ with time and  introduced equations of motion in the projective plane
\bq\label{eq-lag}
 \dfrac{dx_1}{dt}=y_1\,,\quad \dfrac{dy_1}{dt}=\dfrac{3x_1+a}2\qquad\mbox{and}\qquad
    \dfrac{dx_2}{dt}=y_2\,,\quad \dfrac{dy_2}{dt}=\dfrac{3x_2+a}{2}\,,
\eq
 associated with  differential $\omega=dx/y$ of the first kind on  $X$.
 These equations of motion in the projective plane have an integral of motion, i.e. fixed point  $P_3$  in  Fig.1.  All   details of the Lagrange calculations can be found on  page 144 of Greenhill's textbook \cite{gr} and in \cite{grif04}.

In  Clebsch and Gordan's interpretation of Abel's result   points $P_1,P_2$ and $P_3$  form an intersection divisor of plane curves $X$ and  $Y$
\bq\label{eq-add1}
D=P_1+P_2+P_3=0\,,
\eq
where $+$ and $=$ are addition and linear equivalence of divisors on $X$.  There are two well-known interpretations of   intersection divisor $D$:
\begin{enumerate}
  \item $P_1$ and $P_2$ form an effective divisor or the point of $X (2)$, whereas  $P_3$ is a point of $X(1)$. In this case equation (\ref{eq-add1}) and Fig.1. describe canonical injection $j_{12}: X(2)\to X(1)$ so that
      \[j_{12}(P_1,P_2)= P_3\,.\]
  \item $P_1$, $P_2$ and $P_3$ are elements of the Jacobian of $X$. In this case equation (\ref{eq-add1}) and   Fig.1. describe a  group law of algebraic group $Jac(X)$
  \[P_1+P_3=-P_2\,.\]
\end{enumerate}
According to  Jacobi we can identify $X(k)$  with Lagrangian submanifolds  in  phase space $M\simeq \mathbb R^{2k}$ with respect to a family of compatible Poisson brackets \cite{jac36}. In this case two mathematical  interpretations of   intersection divisor $D$ generates two physical interpretations of the corresponding  Poisson maps:
\begin{enumerate}
  \item $P_1$ and $P_2$ describe evolution of some dynamical system with two degrees of freedom with respect to time $t$, whereas coordinates of  $P_3$ are  integrals of motion (superintegrable systems);
  \item $P_1$, $P_2$ are states of some dynamical system with one degree of freedom at $t$ and $t+\Delta t$, whereas fixed point $P_3$ plays the role of  discretization step (integrable discrete maps).
\end{enumerate}
The first interpretation appeared in the Euler and Lagrange investigations of the two-center problem. The second interpretation is closely related with  so-called B\"cklund transformations of the Hamilton-Jacoby equation.
\end{exam}

There is one algebraic group $Jac(X)$ associated with  curve $X$ and only two families of Poisson maps generated by
addition and multiplication on the Jacobian.  Canonical injections
\[
j_{mn}:\qquad X(n)\to X(m)\,,\qquad n>m\,.
\]
 generate many other Poisson maps which properties  have not been studied till now. Nevertheless, we have some examples of application of these maps for studying relations between various integrable system  \cite{ts15a,ts15b,ts15c} and constructing new integrable systems \cite{ts17p,ts17v,ts17c,ts17e}.

 There are also other  relations  between  symmetric products $X(k)$  of  algebraic curve $X$
 \[
\tilde{j}_{mn}:\qquad X(n)\to X(m)
\]
 without restriction $n>m$,  which  can be used in classical mechanics.

\begin{exam}
Let us take a family of parabolas $Y'$
\[
 Y':\qquad y-c=x\Bigl(b_1(t)x+b_0(t)\Bigr)\,.
\]
all passing through the same fixed point and depending  on  parameter $t$, see   Fig.2.
   \begin{figure}[!ht]
\center{\includegraphics[width=0.7\linewidth, height=0.2\textheight]{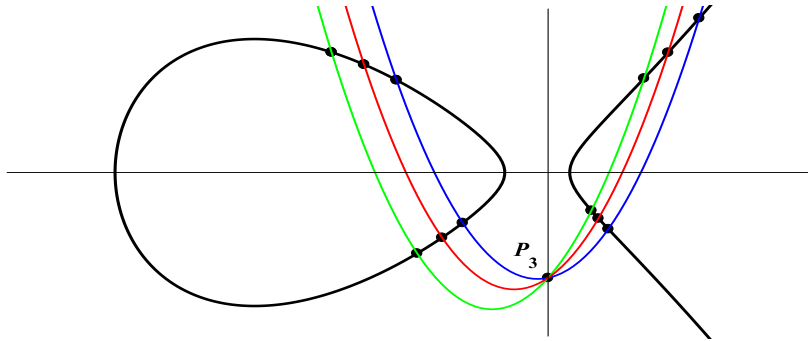} }
\caption{Motion points $P_1,P_2$ and $P_4,P_5$ around  fixed point $P_3$}
\end{figure}
The equation for $Y'$ is obtained by multiplying the
equations for  straight line $Y$ by $x$ and shifting the ordinate by an arbitrary
parameter $c$.

At any $t$ five  points $P_1,\ldots,P_5$  lie on  parabola $Y'$ and movable points satisfy
\bq\label{eq-add2}
D'=P_1+P_2+P_4+P_5=0\,,\qquad D'\in\mbox{Div}X'\,.
\eq
Fixed point $P_3$ does not belong to $X$ and, therefore,  equation (\ref{eq-add2})  does not include this point in contrast with equation (\ref{eq-add1}).   This equation and Fig.2. determine a  mapping $\tilde{j}_{22}:X(2)\to X(2)$ so that
\[
j_{22}(P_1,P_2)\to (P_4,P_5)\,.
\]
 In \cite{ts17c}  we applied this map to construction of a  new integrable system on a plane with two integrals of motion which are polynomials of second and six order in momenta.
\end{exam}

In this note we continue to discuss applications of  equations (\ref{eq-add1}) and (\ref{eq-add2}) in classical mechanics.
Our  main aim is to draw attention to the possibilities of using well-known and no-so-well-known relations between symmetric products   of the algebraic curves in classical mechanics that are not inferior to the possibilities of using group operations on Jacobian, torsion subgroup actions on Jacobian, isogenies of Jacobians, etc.   All the examples below will be related to cubic curve $X$ in  the Weierstrass form (\ref{wei-eq}) in order to discuss the most simplest  integrable systems.

\section{Abel's sums with holomorphic differentials}
Let us rewrite equation (\ref{eq-add1}) in its  expanded form.  At any time coordinates of  two points $P_i$ and $P_j$  determine coefficients
 \[
b_1(t)=\dfrac{y_i-y_j}{x_i-x_j}\,,\qquad b_0(t) =\dfrac{x_iy_j-x_jy_i}{x_i-x_j}
\]
and  coordinates of  third point $P_k$
\bq\label{add-w}
x_k=b_1^2(t)-(x_i+x_j)\,,\qquad y_k=b_1(t)x_k+b_0(t)\,.
\eq
In classical mechanics coordinates of movable points $P_1=(x_1,y_1)$ and $P_2=(x_2,y_2)$  can be identified:
\begin{itemize}
  \item for dynamical system with one degree of freedom   with coordinates $(q,p)$ on  phase space $M$ at  two different times
\bq\label{1-xy}
x_1=q(t_n)\,,\quad y_1=p(t_n)\qquad\mbox{and}\qquad
x_2=q(t_{n+1})\,,\quad y_1=p(t_{n+1})\,.
\eq
  \item for dynamical system with two degrees of freedom  with coordinates $(u_1,p_{u_1})$ and $(u_2,p_{u_2})$ on  phase space $M$ at the same time
  \bq\label{2-xy}
x_1=u_1(t)\,,\quad y_1=p_{u_1}(t)\qquad\mbox{and}\qquad
x_2=u_2(t),,\quad y_1=p_{u_2}(t)\,.
\eq
\end{itemize}
In the first case we rewrite equation (\ref{eq-add1}) in the form
\[
P_2(t_{n+1})=-P_1(t_n)-P_3
\]
and interpret it as a discrete map $P(t_{n})\to P(t_{n+1})$ depending on some fixed parameter $P_3$.

In the second  case we rewrite equation (\ref{eq-add1}) in the form
\[
P_3=-P_1(t)-P_2(t)
\]
and interpret it as a definition of the additional first integral $P_3$.

Thus, discretization and superintegrability have all combined  into one arithmetic equation in Div$X$.

\subsection{ Integrable discrete map}
Let us identify  $X$ with Lagrangian submanifold in phase space  $M=T^*\mathbb R$ and
consider Lagrange equation of motion (\ref{eq-lag}) on a cubic curve (\ref{wei-eq})
\[
\dfrac{dq}{dt}=p\qquad\mbox{where}\qquad p^2=q^3+aq+b\,.
\]
This equation appears when we take  the Hamilton function
\bq\label{1-ham}
H={p^2}-q^3-aq\,,\qquad
\eq
and canonical Poisson brackets $\{q,p\}=1$, which define Hamiltonian equations of motion
\bq\label{1-eq}
\dot{q}=\dfrac{\partial H}{\partial p}=2p\,,\qquad \dot{p}=-\dfrac{\partial H}{\partial q}=3q^2+a\,.
\eq
At $H=b$ these equations are reduced to  (\ref{eq-lag}).

The same Lagrange equation (\ref{eq-lag}) appears when we consider the motion of the symmetric heavy top. In the Lagrange case
equation for  nutation  $\gamma_3(t)$
\bq\label{1-eq2}
\left(\dfrac{d\gamma_3}{dt}\right)^2=(1-\gamma_3)(h-k^2-2\gamma_3)-(c-k \gamma_3)^2
\eq
is also reduced to  (\ref{eq-lag}), see\cite{lag} and \cite{jac69}. Here $c, h$ and $k$ are the values of the corresponding integrals of motion.

According  \cite{ts18a,ts18b} equation (\ref{eq-add1}) is a finite-difference equation, which determine exact two-point discretization
of equations of motion (\ref{1-eq}) or (\ref{1-eq2}). Indeed, substituting (\ref{1-xy}) and $i=1$, $j=3$, $k=2$ into  (\ref{add-w}) one gets
\[
q(t_{n+1})=b_1^2-\Bigl(q(t_n) +x_3\Bigr)\,,\qquad p(t_{n+1})=p(t_n)+b_1\Bigl(x_3-q(t_n)\Bigr)\,,\qquad
                                                                            b_1=\frac{y_3-p(t_n)}{x_3-q(t_n)}\,.
\]
If we also put in $x_3=\lambda_n$ and $y_3=\sqrt{\lambda_n^3+a\lambda_n+b}$, we obtain an iterative system of 2-point invertible mappings $M\to M$ depending on a family of parameters $\lambda_k$
\vskip0.1truecm
\[
\cdots\,\xrightarrow[\lambda_{k-2}]{}\left(
  \begin{array}{c}
    q(t_{k-1}) \\
    p(t_{k-1}) \\
  \end{array}\right)\xrightarrow[\lambda_{k-1}]{}\left(
  \begin{array}{c}
    q(t_{k}) \\
    p(t_{k})\\
  \end{array}
\right)
\xrightarrow[\lambda_{k}]{}\left(
  \begin{array}{c}
    q(t_{k+1}) \\
    p(t_{k+1}) \\
  \end{array}
\right)\xrightarrow[\lambda_{k+1}]{}\,\cdots.
\]
\vskip0.1truecm
\par\noindent
It is the so-called exact discretization of the equations of motion (\ref{1-eq}) preserving the form of integrals of motion and  Poisson bracket, see e.g. \cite{ts17v,ts17c,ts18a,ts18b,ts18d}.
\begin{prop}
Equation (\ref{eq-add1}) in Div$X$ yields an integrable discrete map on phase space $M\to M$, dim$M=1$, preserving the form of integrals of motion and the canonical Poisson bracket.
\end{prop}
Here we  explicitly present the exact discretization of  motion in cubic potential and of motion of the Lagrange top associated with elliptic curve in the Weierstrass form (\ref{wei-eq}). In similar way we can take an elliptic curve in the Jacobi form
\[
X:\quad y^2=ax^4+bx^2+c
\]
and obtain  exact discretization of the Duffing oscillator \cite{ts18a,ts18b} and of the Euler top \cite{ts18bb}.

\subsection{Superintegrable system with two degrees of freedom}
Let symmetric product $X(2)$ be a Lagrangian submanifold in phase space  $M=T^*\mathbb R^2$.
If we identify abscissas and ordinates of  points $P_1$ and $P_2$ with canonical coordinates (\ref{2-xy}) on  phase space $M$  and solve a pair of equations (\ref{wei-eq})
\bq\label{sep-rel11}
\begin{array}{rcl}
 p_{u_1}^2&=&u_1^3+au_1+b\,,\\ \\
 p_{u_2}^2&=&u_2^3+au_2+b
 \end{array}
\eq
with respect to  $a$ and $b$, we obtain two functions on the phase space
\bq\label{wei-int}
\begin{array}{rcl}
a&=&\dfrac{p_{u_1}^2}{u_1-u_2}+\dfrac{p_{u_2}^2}{u_2-u_1}-u_1^2-u_1u_2-u_2^2\,,\\
\\
b&=&\dfrac{u_2p_{u_1}^2}{u_2-u_1}+\dfrac{u_1p_{u_2^2}}{u_1-u_2}+(u_1+u_2)u_1u_2\,,
\end{array}
\eq
which are in involution with respect to the canonical Poisson bracket
\[
\{u_1,u_2\}=0\,,\quad \{p_{u_1},p_{u_2}\}=0\,,\quad
\{u_i,p_{u_j}\}=\delta_{ij}\,,
\]
Taking  $H=a$ as a Hamiltonian, one gets integrable system on phase space $M=T^*\mathbb R^2$
with Hamiltonian equations of motion
\bq\label{2-eq}
\dot{u}_i=\dfrac{\partial H}{\partial p_{u_i}}\,,\qquad \dot{p}_{u_i}=-\dfrac{\partial H}{\partial u_i}\,,
\eq
which are reduced to quadratures
\bq\label{q-1}
\int \dfrac{u_1du_1}{\sqrt{u_1^3+au_1+b}}+
\int \dfrac{u_2du_2}{\sqrt{u_2^3+au_2+b}}=-2t
\eq
and
\bq\label{q-2}
\int \dfrac{du_1}{\sqrt{u_1^3+au_1+b}}+
\int \dfrac{du_2}{\sqrt{u_2^3+au_2+b}}=\mathrm{const}\,.
\eq
According Euler and Lagrange \cite{eul1,lag}, the   first quadrature determines parameterization of trajectories, whereas the second quadrature determines the form of trajectories. Thus, we can use first quadrature for discretization of time variable and
second quadrature to the search of algebraic trajectories associated with additional algebraic integral of motion.

We identify second quadrature (\ref{q-2}) and the corresponding Abel's sum with equation (\ref{eq-add1})
\[
P_1(t)+P_2(t)=P_3=\mathrm{const}\,.
\]
Substituting (\ref{2-xy}) and $i=1$, $j=2$, $k=3$ into (\ref{add-w})   one gets
additional first integrals of equations of motion (\ref{2-eq})
\bq\label{2-add-int}
x_3=\left(\dfrac{p_{u_1}-p_{u_2}}{u_1-u_2}\right)^2-u_1-u_2\,,\qquad
y_3=p_{u_1}+\left(\dfrac{p_{u_1}-p_{u_2}}{u_1-u_2}\right)(x_3-u_1)\,.
\eq
Functions $a, b$  (\ref{wei-int}) and functions $x_3, y_3$ (\ref{2-add-int}) on  phase space  $M=T^*\mathbb R^2$ form an algebra of integrals
\bq\label{alg-int11}
\begin{array}{lll}
 \{a,b\}=0\,,\qquad &\{a,x_3\}=0\,,\qquad &\{a,y_3\}=0\,,\\ \\
 \{b,x_3\}=2y_3\,,\qquad &\{b,y_3\}=3x_3^2+a\,,\qquad &\{x_3,y_3\}=-1\,,
 \end{array}
\eq
in which Weierstrass equation (\ref{wei-eq}) plays the role of syzygy
\[
y_3^2=x_3^3+ax_3+b\,.
\]
\begin{prop}
Equation (\ref{eq-add1}) in Div$X$ describes a representation of the algebra of integrals (\ref{alg-int11}), i.e. superintegrable system
on phase space $M$, dim$M=2$.
\end{prop}
First equation in  (\ref{2-add-int}) is nothing more than an additional law for the Weierstrass function
\[
\wp(u_1+u_2)=\dfrac{1}{4}\left(\dfrac{\wp'(u_1)-\wp'(u_2)}{\wp(u_1)-\wp(u_2)}\right)^2-\wp(u_1)-\wp(u_2)\,,
\]
and, therefore,  algebra of the first integrals coincides with well-know relations between Weierstrass $\wp$-function and its derivatives, see \cite{cal,gr,hens}.

After canonical transformation of variables
\[u_1 = q_1-\sqrt{q_2},\quad u_2 = q_1+\sqrt{q_2},\quad
p_{u_1}=\dfrac{p_1}{2}+\dfrac{(u_1-u_2)p_2}{2},\quad p_{u_2}=\dfrac{p_1}{2}-\dfrac{(u_1-u_2)p_2}{2}\]
these first integrals  look like
\[\begin{array}{ll}
a=H=p_1p_2-3q_1^2-q_2\,,\qquad &b=
\dfrac{p_1^2}{4}-q_1p_1p_2+q_2p_2^2+2q_1(q_1^2-q_2)\,,\\ \\
x_3=p_2^2-2q_1\,,\qquad
&y_3=\dfrac{p_1}2+p_2^3-3p_2q_1\,.
\end{array}
\]
Similar superintegrable systems on the plane with quadratic Hamiltonians
\[H=p_1p_2+V(q_1,q_2)\]
and cubic first integrals are discussed in  \cite{ts11,ts08a,ts09}.

 \subsection{Other representations of the algebra of first integrals}
Below we consider the arithmetic equation
\[
D=\sum_{i=1}^k n_iP_i=0\,,\qquad n_i\in\mathbb Z
\]
and  Abel's sum with holomorphic differentials involving more than three terms
\[
\sum_{i=1}^k  n_i\int \omega(x_i,y_i)dx_i=const\,.
\]
 Different exact discretizations  of Hamiltonian and non-Hamiltonian equations of motion
  associated with such arithmetic equations in Div$X$ are discussed in \cite{ts17v,ts17c,ts18a,ts18b,ts18bb,ts18d}.

Superintegrable systems  associated with the same arithmetic equations  are discussed in \cite{ts18c,ts18s}.
These superintegrable systems can be considered as different representations of the algebra of integrals (\ref{alg-int11})
\[
\begin{array}{lll}
 \{a,b\}=0\,,\qquad &\{a,x_3\}=0\,,\qquad &\{a,y_3\}=0\,,\\ \\
 \{b,x_3\}=2y_3\,,\qquad &\{b,y_3\}=f'(x_3)\,,\qquad &\{x_3,y_3\}=-1\,,
 \end{array}
\]
 labelled by two integers $n$ and $m$. Here $f'$ is a derivative of function $f$ from the definition of  elliptic curve $X:\,y^2=f(x)$.
Indeed,  let us make a trivial non-canonical transformation
\[
p_{u_1}\to\dfrac{p_{u_1}}{n}\,,\qquad p_{u_2}\to\dfrac{p_{u_2}}{m}
\]
in the separated relations (\ref{sep-rel11}) \cite{ts18c}.  Solving the new separated relations
\bq\label{sep-relNM}
\begin{array}{rcl}
\left(\dfrac{p_{u_1}}{n}\right)^2&=&u_1^3+au_1+b\,,\\ \\
 \left(\dfrac{p_{u_2}}{m}\right)^2&=&u_2^3+au_2+b
 \end{array}
\eq
with respect to  $a$ and $b$, we obtain two functions on the phase space
\bq\label{wei-intNM}
\begin{array}{rcl}
a&=&\dfrac{p_{u_1}^2}{n^2(u_1-u_2)}+\dfrac{p_{u_2}^2}{m^2(u_2-u_1)}-u_1^2-u_1u_2-u_2^2\,,\\
\\
b&=&\dfrac{u_2p_{u_1}^2}{n^2(u_2-u_1)}+\dfrac{u_1p_{u_2^2}}{m^2(u_1-u_2)}+(u_1+u_2)u_1u_2\,,
\end{array}
\eq
which are in  involution with respect to the canonical Poisson bracket. Taking  $H=a$ as a Hamiltonian, one gets  Hamiltonian equations of motion (\ref{2-eq} )
\bq\label{eq-mNM}
\dfrac{du_1}{dt}=\dfrac{2p_{u_1}}{n^2(u_1-u_2)}\qquad\mbox{and}\qquad \dfrac{du_2}{dt}=\dfrac{2p_{u_2}}{m^2(u_2-u_1)}\,.
\eq
Substituting $p_{u_1}$ and $p_{u_2}$ from (\ref{sep-relNM})  into (\ref{eq-mNM}) we obtain equations
\[
\dfrac{nd u_1}{\sqrt{u_1^3+au_1+b}}=\dfrac{2dt}{u_1-u_2}\qquad\mbox{and}\qquad
\dfrac{md u_2}{\sqrt{u_2^3+au_2+b}}=\dfrac{2dt}{u_2-u_1}\,.
\]
which are reduced to quadratures
\[
n\int \dfrac{u_1du_1}{\sqrt{u_1^3+au_1+b}}+
m\int \dfrac{u_2du_2}{\sqrt{u_2^3+au_2+b}}=-2t
\]
and
\[
n\int \dfrac{du_1}{\sqrt{u_1^3+au_1+b}}+
m\int \dfrac{du_2}{\sqrt{u_2^3+au_2+b}}=\mathrm{const}\,.
\]
Second quadrature is the homogeneous Abel's sum associated with an arithmetic equation in Div$X$
\bq \label{add-eq1-mn}
[n]P_1+[m]P_2+P_3=0\,.
\eq
Coordinates of the  fixed point $P_3=(x_3,y_3)$
\bq\label{eul-int-nm}
 x_3=\left(\dfrac{[n]y_1-[m]y_2}{[n]x_1-[m]x_2} \right)^2-\bigl([n]x_1+[m]x_2\bigr)\,,\qquad
 y_3^2=x_3^3+ax_3+b
 \eq
are functions on the phase space commuting with Hamiltonian $H=a$ (\ref{wei-intNM}). Here $[k]x$ and $[k]y$ are affine coordinates of point $[k]P$ on the projective plane defined by  well-known  equation
 \[
 [k]P\equiv\bigl([k]x,[k]y\bigr)=\left(x-\dfrac{\psi_{k-1}\psi_{k+1}}{\psi_k^2}\,,\dfrac{\psi_{2k}}{2\psi_k^4} \right)
 \]
 where $\psi_k(x,y)$ are the so-called division or torsion polynomials in a ring $\mathbb Z[x,y,a,b]$, see  \cite{lang78,was08}. The first four polynomials are defined explicitly as
 \[\begin{array}{rcl}
 \psi_1&=&1,\quad \psi_2=2y\,,\quad \psi_3=3x^4+6ax^2+12bx-a^2\,,\\
 \\
 \psi_4&=&4y(x^6+5ax^4+20bx^3-5a^2x^2-4abx-8b^2-a^3)\,,
 \end{array}
  \]
 the subsequent polynomials are defined inductively as
  \[\begin{array}{rcll}
\psi_{2k+1}&=&\psi_{k+1}\psi_k^3-\psi_{k-1}\psi^3_{k+1}\,,\qquad &k\geq 2\,,\\ \\
\psi_{2k}&=&(2y)^{-1}\psi_k(\psi_{k+1}\psi^2_{k-1}-\psi_{k-2}\psi^2_{k+1}\,,\qquad &k\geq 3 \,.
 \end{array}
  \]
Using these division polynomials we can easily calculate integrals of motion (\ref{eul-int-nm}). For instance, at $n=2$ and $m=1$  additional first integral $x_3$ is a rational function of the form
\[\begin{array}{rcl}
x_3&=&4   u_1-3   u_2-\dfrac{8   (u_1-u_2)\left(6   u_1^3-12   u_1^2 u_2+6   u_1 u_2^2+p_{u_1} p_{u_2}-2 p_{u_2}^2\right)}{\left(8   u_1^3-12   u_1^2 u_2+4   u_2^3-p_{u_1}^2+4 p_{u_1} p_{u_2}-4 p_{u_2}^2\right)}\\
\\
&+&\dfrac{64  ^2 (u_1-u_2)^4\left(2   u_1^3-3   u_1^2 u_2+  u_2^3+p_{u_1} p_{u_2}-p_{u_2}^2\right)}{\left(8   u_1^3-12   u_1^2 u_2+4   u_2^3-p_{u_1}^2+4 p_{u_1} p_{u_2}-4 p_{u_2}^2\right)^2}\,.
\end{array}
\]
At $n=3$ and $m=1$ the additional first integral is equal to
\[
x_3=-(u_1+u_2)+\dfrac{(p_{u_1}^2-9p_{u_2}^2)^2}{81(p_{u_1}+p_{u_2})^2(u_1-u_2)^2}+
\dfrac{8p_{u_1}^2A_1}{9(p_{u_1}+p_{u_2})^2B}-\dfrac{64p_1^5A_2}{9(p_{u_1}+p_{u_2})B^2}
\]
where
\[
B=p_{u_1}^4-8 p_{u_1}^3 p_{u_2}+18\left( p_{u_2}^2- u_1^2 u_2+2 u_1 u_2^2- u_2^3\right) p_{u_1}^2-27\left(p_{u_2}^2-2 u_1^3+3 u_1^2 u_2-u_2^3\right)^2\,,
\]
and
\[\begin{array}{rcl}
A_1&=&
(15 u_1+19 u_2) p_{u_1}^4-6(13 u_1+5 u_2) p_{u_1}^3 p_{u_2} -3 (11 u_1+19 u_2) (2 u_1+u_2) (u_1-u_2)^2 p_{u_1}^2\\ \\
&& +54\bigl((u_1+u_2)p_{u_2}^2+(2u_1+u_2)(u_1-u_2)^3\bigr) p_{u_1}p_{u_2} \\ \\
&&-27 p_{u_2}^2 (5 u_1+u_2) \bigl( p_{u_2}^2-(2u_1+u_2)(u_1-u_2)^2 \bigr)\,,
\\ \\
A_2&=&
2(u_1+u_2) p_{u_1}^4-4 (7 u_1+5 u_2) p_{u_1}^3 p_{u_2}+54 p_{u_2}^2 (5 u_1+u_2)\bigl (p_{u_2}^2-(2u_1+u_2)(u_1-u_2)^2\bigr)\\ \\
&&+3\bigl(24(2 u_1+ u_2) p_{u_2}^2-(7 u_1^2+16 u_1 u_2+13 u_2^2) (u_1-u_2)^2\bigr) p_{u_1}^2\\
\\
&& -9 \bigl(12(3u_1+u_2)p_{u_2}^2-(23u_1^2+38u_1u_2+11u_2^2)(u_1-u_2)^2\bigr) p_{u_1}p_{u_2}\,.
\end{array}
\]
In both cases of $m=2$ and $m=3$  four integrals of motion $a,b$ (\ref{wei-intNM})  and  $x_3,y_3$ (\ref{eul-int-nm}) form  the algebra of integrals (\ref{alg-int11}), same as in the previous case at $n=m=1$.

We conjecture this is to be true for general  $n$ and $m$.
\begin{prop}
Equation (\ref{add-eq1-mn}) in Div$X$ describes  representation of the algebra of integrals (\ref{alg-int11}) labelled by two integers $n$ and $m$, i.e.  superintegrable system on  phase space $M$, dim$M=2$.
\end{prop}
Other examples of superintegrable systems associated with  Abel's sums  including holomorphic differentials may be found in \cite{ts08a,ts09,ts10, ts18c,ts18s}.

\section{Abel's sums with non-holomorphic differentials}
Let us consider the motion of parabola $Y'$ defined by an equation of the form
\[
 Y':\qquad y-c=x\bigl(b_1(t)x+b_0(t)\bigr)\,.
\]
around  fixed point $P_3=(0,c)$, see Fig.2. If $P_i(t)$ and $P_j(t)$ are two movable intersection points of parabola $Y'$ with cubic curve $X$ (\ref{wei-eq}), then
\bq
\label{coeffs-b2}
b_1(t)=\dfrac{y_i  x_j-y_j  x_i}{x_j x_i  (x_i -x_j)}\qquad\mbox{and}\qquad
b_0(t)=-\dfrac{y_i  x_j^2-y_j  x_i ^2}{x_j x_i  (x_i -x_j)}
\eq
due to Lagrange interpolation of parabola  using three points $P_i(t),P_j(t)$ and $P_3$.

Equation (\ref{eq-add2})
\[
D'=P_1+P_2+P_4+P_5=0\,,\qquad D'\in\mbox{Div}X\,,
\]
can be considered as a discrete map in Div$X$
\[(P_i,P_j)\to (P_\ell,P_m) \]
because  coordinates of the remaining two movable  points $P_\ell$ and $P_m$ are easily  expressed via $x_i,x_j$ and $y_i,y_j$. Indeed, according to Abel \cite{ab}  abscissas $x_\ell$ and $x_m$ are  roots of the so-called Abel polynomial
\bq\label{ab-pol}
\Psi=x^3+ax+b-\Bigl(x\bigl(b_1x+b_0)+c\Bigr)^2=b_1^2(x-x_i)(x-x_j)(x-x_\ell)(x-x_m)\,,
\eq
whereas ordinates $y_\ell$ and $y_m$ are equal to
\bq\label{ab-y}
y_\ell=x_\ell\bigl(b_1x_\ell+b_0)+c\,,\qquad y_m=x_m\bigl(b_1x_m+b_0)+c\,.
\eq
As mentioned above, discrete map in Div$X$ generates an integrable discrete map on phase space $M$.

\subsection{Integrable discrete map}
Let us come back to the integrable system with  two degrees of freedom defined  by the following integrals of motion (\ref{wei-int})
\[
\begin{array}{rcl}
a&=&\dfrac{p_{u_1}^2}{u_1-u_2}+\dfrac{p_{u_2}^2}{u_2-u_1}-u_1^2-u_1u_2-u_2^2\,,\\
\\
b&=&\dfrac{u_2p_{u_1}^2}{u_2-u_1}+\dfrac{u_1p_{u_2^2}}{u_1-u_2}+(u_1+u_2)u_1u_2\,,
\end{array}
\]
These integrals are in the involution with respect to the Poisson brackets
\[
\{u_1,u_2\}_\varphi=0\,,\quad \{p_{u_1},p_{u_2}\}_\varphi=0\,,\quad
\{u_i,p_{u_j}\}_\varphi=\delta_{ij} \varphi_i(u_i,p_{u_i})\,,
\]
labelled by two arbitrary functions $\varphi_i(u_i,p_{u_i})$. The corresponding Poisson bivector reads as
\bq\label{poi-f}
\Pi=\left(
      \begin{array}{cccc}
        0 & 0 & \varphi_1(u_1,p_{u_1}) & 0 \\
        0 & 0 & 0 & \varphi_2(u_2,p_{u_2}) \\
        -\varphi_1(u_1,p_{u_1}) & 0 & 0 & 0 \\
        0 & -\varphi_2(u_2,p_{u_2}) & 0 & 0 \\
      \end{array}
    \right)\,.
\eq
Taking  $H=a$ as a Hamiltonian, one gets an integrable system on the phase space $M=T^*\mathbb R^2$
with Hamiltonian equations of motion (\ref{2-eq}) which are reduced to quadratures (\ref{q-1}) and (\ref{q-2}).

In the previous Section we use second quadrature (\ref{q-2}),  i.e. Abel's sum with the holomorphic differential on $X$
 \[
 \int \dfrac{du_1}{\sqrt{u_1^3+au_1+b}}+
\int \dfrac{du_2}{\sqrt{u_2^3+au_2+b}}=\mathrm{const},
\]
 to construct the additional first integrals. These integrals $x_3$ and $y_3$ coincide with  coordinates of the fixed point $P_3$ on  $X$, the existence of additional algebraic integral of motion guarantees an existence of algebraic trajectories, see Euler paper \cite{eul1}.

In order to construct a discrete integrable map $M\to M$ we have to take first quadrature (\ref{q-1}), i.e. Abel's sum with non-holomorphic differentials
\[
\int \dfrac{u_1du_1}{\sqrt{u_1^3+au_1+b}}+
\int \dfrac{u_2du_2}{\sqrt{u_2^3+au_2+b}}=-2t\,,
\]
 and interpret equation (\ref{eq-add2})
 \[
 P_1+P_2+P_3+P_4=0
 \]
  as a discrete map relating pairs of  movable points at $t_n$ and $t_{n+1}$, respectively:
  \[(P_i,P_j)(t_n)\to (P_\ell,P_m)(t_{n+1})\,.\]
 For brevity,  we omit dependence on time below.

 Thus, let us identify variables on  phase space $M$
with affine coordinates of two movable  points  $P_1$ and $P_2$ on a projective plane
\[
x_1=u_1\,,\quad y_1=p_{u_1}\qquad\mbox{and}\qquad
x_2=u_2\,,\quad y_1=p_{u_2}\,.
\]
Coordinates of remaining  movable  points  $P_4$ and $P_5$ are some other variables on  $M$
\[
x_4=v_1\,,\quad -y_4=p_{v_1}\qquad\mbox{and}\qquad
x_5=v_2\,,\quad -y_5=p_{v_2}\,.
\]
Here we change sign before ordinates $y_{4,5}$ in order to rewrite the equation (\ref{eq-add2}) in the following form
\[P_1+P_2=P_4+P_5\,.\]
At $c=0$ in (\ref{ab-pol}) and (\ref{ab-y}) one gets
\[\begin{array}{rcl}
(x-v_1)(x-v_2)=x^2&+&
\dfrac{(u_1-u_2) \bigl(p_{u_1}^2 u_2^2-p_{u_2}^2 u_1^2-(u_1-u_2)u_1^2u_2^2\bigr)}{(p_{u_1} u_2-p_{u_2} u_1)^2}\,x\\ \\
&+&\dfrac{u_1 u_2 (u_1-u_2) (p_{u_1}^2 u_2-p_{u_2}^2 u_1-u_1^3 u_2+u_1 u_2^3)}{(p_{u_1} u_2-p_{u_2} u_1)^2}
\end{array}
\]
and
\[
p_{v_1}=-v_1\bigl(b_1v_1+b_0)\,,\qquad p_{v_2}=-v_2\bigl(b_1v_2+b_0)\,,
\]
where
\[
b_1=\dfrac{p_{u_1} u_2-p_{u_2} u_1}{u_1 u_2 (u_1-u_2)}\,,\qquad
b_0=-\dfrac{(p_{u_1} u_2^2-p_{u_2} u_1^2)}{u_1 u_2 (u_1-u_2)}\,.
\]
 Thus, we have an integrable discrete map on the phase space $M$
 \[
 \rho:\quad (u_1,u_2,p_{u_1},p_{u_2})\to (v_1,v_2,p_{v_1},p_{v_2})
  \]
 preserving form of the  integrals of motion (\ref{wei-int}) and form of the following  Poisson bivector
 \[\rho:\quad
 \Pi=\left(
      \begin{array}{cccc}
        0 & 0 &u_1 & 0 \\
        0 & 0 & 0 & u_2 \\
        -u_1 & 0 & 0 & 0 \\
        0 & -u_2 & 0 & 0 \\
      \end{array}
    \right)
    \to
      \Pi=\left(
      \begin{array}{cccc}
        0 & 0 & v_1 & 0 \\
        0 & 0 & 0 & v_2 \\
        -v_1 & 0 & 0 & 0 \\
        0 & -v_2& 0 & 0 \\
      \end{array}
    \right)\,,
  \]
which belongs to a family of compatible Poisson bivectors (\ref{poi-f}).

\begin{prop}
Equation (\ref{eq-add2}) in Div$X$ yields an integrable discrete map $M\to M$, dim$M=2$, preserving the form of integrals of motion and one of the compatible Poisson bivectors  (\ref{poi-f}).
\end{prop}

Other examples of the Poisson maps associated with differentials $xdx/y$ and $x^2dx/y$ on an elliptic curve  may be found in \cite{ts17p}.

\subsection{Construction of integrable systems with higher order polynomial integrals of motion}
Let us identify symmetric product $X(2)$ with Lagrangian submanifold in phase space  $M=T^*\mathbb R^2$ such that
 affine coordinates of movable  points are expressed via canonical variables on $M$
 in the following way
 \[
x_1=u_1\,,\quad y_1=u_1p_{u_1}\qquad\mbox{and}\qquad
x_2=u_2\,,\quad y_1=u_2p_{u_2}
\]
and
\[
x_4=v_1\,,\quad -y_4=v_1p_{v_1}\qquad\mbox{and}\qquad
x_5=v_2\,,\quad -y_5=v_2p_{v_2}\,.
\]
Then we determine canonical transformation $\rho':M\to M$ preserving standard Poisson bivector
\[\rho':\quad
 \Pi=\left(
      \begin{array}{cccc}
        0 & 0 &1 & 0 \\
        0 & 0 & 0 & 1 \\
        -1 & 0 & 0 & 0 \\
        0 & -1 & 0 & 0 \\
      \end{array}
    \right)
    \to
      \Pi=\left(
      \begin{array}{cccc}
        0 & 0 & 1 & 0 \\
        0 & 0 & 0 & 1 \\
        -1 & 0 & 0 & 0 \\
        0 & -1& 0 & 0 \\
      \end{array}
    \right)\,.
  \]
This canonical transformation is defined by the same relations (\ref{coeffs-b2}), (\ref{ab-pol}) and (\ref{ab-y}).

Usually coordinates $u_1,u_2$ are standard curvilinear orthogonal coordinates on the plane, sphere or ellipsoid, whereas  new canonical coordinates $v_{1,2}$ and $p_{v_{1,2}}$ are images of the curvilinear coordinates after some integrable Poisson maps. In \cite{ts17p,ts17v,ts17c,ts17e,ts18d} we used these coordinates to construct new integrable systems in the framework of the Jacobi method.

For instance, let us take coordinates $v_{1,2}$ and momenta $p_{v_{1,2}}$ defined by (\ref{ab-pol}-\ref{ab-y})
\[\begin{array}{rcl}
(x-v_1)(x-v_2)&=&x^2+\frac{(u_1-u_2)(p_{u_1}^2-p_{u_2}^2-u_1+u_2)}{(p_{u_1}-p_{u_2})^2}\,x
+\frac{(u_1-u_2)(p_{u_1}^2u_1-p_{u_2}^2u_2-u_1^2+u_2^2)}{(p_{u_1}-p_{u_2})^2}\,,\\
\\
v_jp_{v_j}&=&-\dfrac{v_j\bigl(v_j(p_{u_1}-p_{u_2})-p_{u_1}u_2+p_{u_2}u_1\bigr)}{u_1-u_2}\,, \qquad j=1,2.
\end{array}
\]
Substituting these variables into the separated relations
\[
2v_1p_{v_1}^2 =2v_1^2+H+\sqrt{K}\,,\qquad\mbox{and}\qquad
2v_2p_{v_2}^2 =2v_2^2+H-\sqrt{K}\,,
\]
one gets integrable systems with polynomial integrals of motion
\[
H=T+V=\dfrac{u_1(2u_1+u_2)p_{u_1}^2}{u_1-u_2}+\dfrac{u_2(u_1+2u_2)p_{u_2}^2}{u_2-u_1}-2u_1^2-3u_1u_2-2u_2^2\,,
\]
and
\[\begin{array}{rcl}
K&=&\frac{u_1^2u_2^2}{(u_2-u_1)^3}\Bigl(
(3 u_1+u_2) p_{u_1}^4-(u_1+3 u_2) p_{u_2}^4-8  u_1 p_{u_1}^3 p_{u_2}+8u_2p_{u_1}p_{u_2}^3
+6(u_1-u_2)p_{u_1}^2p_{u_2}^2\Bigr.\\
\\
&-&
\Bigl.2 (u_1-u_2) (u_1+3 u_2)p_{u_1}^2+8(u_1^2-u_2^2) p_{u_1}p_{u_2}-2(u_1-u_2)(3u_1+u_2) p_{u_2}^2-(u_1-u_2)^3
\Bigr)\,,
\end{array}
\]
which are polynomials of second  and fourth order in momenta. It is easy to prove that this Hamiltonian has no integrals of motion which are polynomials of first, second and third order in momenta.

Integrable metric
\[
\mathrm g=\left(
            \begin{array}{cc}
             \dfrac{u_1(2u_1+u_2)}{u_1-u_2} & 0 \\
              0 & \dfrac{u_2(u_1+2u_2)}{u_2-u_1} \\
            \end{array}
          \right)
\]
belongs to a family of integrable and superintegrable metrics from \cite{ts17p}. Here we add new potential  $V$ to the known kinetic energy $T$.

Thus, we obtain a new non-trivial integrable  system on a plane with natural quadratic Hamiltonian $H=T+V$  and quartic second integral of motion. This systems belongs to a family of two-dimensional integrable systems with  position-dependent mass (PDM), which has various applications in physics,  see   \cite{ran16,ser19} and references within. Using the proposed approach we can construct other new PDM systems with integrals of motion which are polynomials of second, third, fourth and even sixth order in momenta.

\section{Conclusion}
In modern textbooks Abel's theorem provides the necessary and sufficient conditions for the existence of meromorphic functions with
prescribed zeros and poles on a compact Riemann surface $X$. As is well known, this problem is equivalent to  existence of a parallel section for some complex connection in the holomorphic line bundle of the divisor. It is far from anything in Abel's original  works, although we continue to call it the Abel theorem.

If we come back to original Abel's theorem we can find many applications of this theorem in physics. Indeed, many equations of mathematical physics are reduced to Abel's  quadratures using orthogonal curvilinear coordinates or more exotic variables, for instance, variables of separation for the Kowalevski top. Starting with these well-known  systems we can get new integrable systems and discrete maps by using the proposed approach.

In classical mechanics there are integrable systems with a  common level set of first integrals, which can be identified  with a generalized Jacobian $Jac(X)$, which is a commutative algebraic group. To study these  systems we can use  various  properties of $Jac(X)$ such as  group operations, torsion subgroup actions, isogenies, etc.

There are also  integrable systems with a common level set of first integrals, which  can be identified with   symmetric products $X(n)$ of curve $X$, which  have no a group structure. Nevertheless,  varieties $X(n)$ and their canonical injections  are classical objects of study in algebraic geometry, and  it would be natural to expect to find various applications of these well-studied algebro-geometric tools  in classical mechanics. Unfortunately, we could not find  such applications in the current literature.  In this note we try to fill this gap starting with the simplest cubic curve, its symmetric products and its Jacobian.

The work was supported by the Russian Foundation for Basic Research (project 18-01-00916).

\end{document}